# Quasilattice-conserved optimization of the atomic structure of decagonal Al-Co-Ni quasicrystals

Xiao-Tian Li, Xiao-Bao Yang, and Yu-Jun Zhao[*]

*Department of Physics and State Key Laboratory of Luminescent Materials and Devices,*

*South China University of Technology, Guangzhou 510640, P. R. China*

[*]Corresponding author. Tel: +86-20-87110426; fax: +86-20-87112837;

E-mail: zhaoyj@scut.edu.cn.

**Abstract:** The detailed atomic structure of quasicrystals has been an open question for decades. Here, we present a quasilattice-conserved optimization method (quasiOPT), with particular quasiperiodic boundary conditions. As the atomic coordinates described by basic cells and quasilattices, we are able to maintain the self-similarity characteristics of qusicrystals with the atomic structure of the boundary region updated timely following the relaxing region. Exemplified with the study of decagonal Al-Co-Ni (d-Al-Co-Ni), we propose a more stable atomic structure model based on Penrose quasilattice and our quasiOPT simulations. In particular, "rectangle-triangle" rules are suggested for the local atomic structures of d-Al-Co-Ni quasicrystals.

Since the discovery of quasicrystals, their detailed atomic structure has attracted extensive attention, for vital significance to the understanding of these novel solids. X-ray diffraction experiments were initially applied to the determination of their structures.[1, 2] It was, however, soon proved not easy to distinguish the local isomorphic quasicrystals[3]. The electron microscopy techniques gradually became crucial in the quandary[4], and several influential atomic models of the well-known $Al_{72}Co_8Ni_{20}$ were proposed based on it[5-7]. Nevertheless, it is still difficult for the electron microscopy to distinguish each atom under the resolution limitation.

On the other hand, first-principles calculations[6], together with inter-atomic potential methods[8-10], have also been applied to the quandary. Although theoretical calculations based on first-principles and empirical potentials can offer some insights into the atomic structure, it is far from conclusive for the detailed atomic positions of quasicrystals because of the complicated structure in addition to potential expensive computational efforts. Moreover, traditional boundary treatments, including cluster model and periodic boundary model, have troubles in dealing with quasicrystals. The former strongly depends on the specific atomic arrangement at boundary, while the latter does not match quasicrystals intrinsically. In fact, the feature of quasicrystal structures often disappears when the structures are optimized with traditional boundary conditions.

In this paper, we propose a quasiperiodic boundary model to deal with quasicrystals based on the tiling model. Coupled with the inter-atomic potentials, a quasilattice-conserved optimization method (quasiOPT) is developed for structure optimization of quasicrystals.

With its application to the decagonal Al-Co-Ni (d-Al-Co-Ni), we present a more stable atomic structure model for the well-known quasicrystals.

Tiling and decoration, which fills the space by two or more elementary building blocks, is one of the most prominent models to describe the structure of quasicrystals. This model can be described by two parts: quasilattice and basic cells. Here quasilattice is a general description of the positions and orientations of all building blocks, while basic cells contain information of the detailed atomic position of every type of building block.[11, 12] To maintain the self-similarity of quasicrystals, it is rational to assume that the internal atomic structures are same for each type of cell. For instance, the atomic positions of a simple two dimensional quasicrystal with fivefold symmetry, as shown in Fig. 1c, can be described with a routine quasilattice (as shown in Fig. 1a) and three types of basic cells, a fat cell, a skinny cell, and a point cell. Following the tiling rule, one can construct the quasicrystal (Fig. 1c) with the fat, skinny, and point cells. Unlike the crystals with translation symmetry, unique cells with fewer dimensions such as the point cell are often required to construct quasicrystals.

There is a normative mathematical description for the atomic structure of quasicrystals, as long as we know the internal atomic coordinates of each type of cell and the tiling rule of the cells to the quasilattice. For a general quasilattice, we need to know the positions and orientations of all basic cells to construct it. Typically, four vectors, together with its type $k$, are required to describe a three dimensional cell $\alpha$, a vector ($\vec{R}_\alpha$) for its origin and its three cell vectors ($\vec{a}_{\alpha 1}$, $\vec{a}_{\alpha 2}, \vec{a}_{\alpha 3}$). In addition, the detailed atomic position of each type of basic cell is necessary, and an internal coordinate $\vec{r}'_{ki}$ describes the position of atom $i$ in the $k$-type cell. Thus we can describe the position of atom $i$ in the cell $\alpha$ of type $k$:

$$\vec{r}_{\alpha i} = \vec{R}_\alpha + \vec{r}_{ki}^{\,'} \cdot T_\alpha \tag{1}$$

$$T_\alpha = \begin{pmatrix} \vec{a}_{\alpha 1} \\ \vec{a}_{\alpha 2} \\ \vec{a}_{\alpha 3} \end{pmatrix} \tag{2}$$

Here $T_\alpha$ is the transformation matrix of basic cell $\alpha$, constructed with its three cell vectors. For the special basic cells, one can regard the length of corresponding cell vectors as zero.

Following the above description of the atomic structure of quasicrystals with quasilattice and basic cells, we are able to maintain the self-similarity characteristics by keeping the quasilattice conserved during the simulations by setting up proper relaxing and boundary regions. One easy way is to choose a complete set of basic cells as the relaxing region, and construct a thick boundary region with their atomic coordinates updated timely following the optimization of the corresponding atoms of the relaxing region under the self-similarity rule.

However, basic cells of a specific type usually have various neighboring cells due to the complexity of quasilattice. When the boundary atomic structures are updated following the relaxing region according to the quasilattice, the quasilattice-induced atomic structures may be unphysical, with neighboring atoms too close or too far. For instance, as shown in Fig. 2a, the boundary region (cyan) of a simple quasicrystal is constructed from the relaxing region (magenta). Obviously, some part of the boundary region is unphysical (see the red atom pairs). It is clear that the appearance of unphysical jammed atom pairs is ascribed to the non-equivalent neighboring cells for the corresponding cells in boundary and relaxing regions.

To avoid the unphysical structures, in principle, the relaxing region should include the basic cells with all the possible local structures, i.e., structurally ergodic for the basic cells. Fortunately, it is feasible to enumerate all the possibilities since a perfect quasilattice usually has finite configurations of local structures. For example, it has been demonstrated that, there are only eight types of vertices (local environments) in the well-known Penrose tiling[13]. If the relaxing region includes the atomic structures around all the eight types of vertices, the quasilattice-induced region would be physical on the local structure level. Therefore, we extend the number of basic cells to a complete set with respect to the local structure (within the nearest neighboring cells). That is, it includes the eight types of vertices of the quasilattice, which can be associated with seven fat and four skinny cells as the new relaxing region (cf. Fig. 2b).

A new issue, however, surfaces as we adopt a complete set of basic cells considering various nearest neighbors. When we update the boundary atomic coordinates, there are alternative choices of cells to follow as the specific cells are not unique in the relaxing region now. For instance, there are seven fat and four skinny cells now in our example (cf. Fig. 2b). Here, we take an average effect for the seven fat and four skinny cells respectively, so all the basic cells of the same type would have the same atomic structure. At first sight, it seems ill-considered to take such an average because their boundary structures are different. In fact, for a good model of quasicrystals, the basic cells usually have similar local environment at the atom level.

One more issue to be addressed is how to deal with the atoms crossing the boundary of a cell during optimization. For crystals, it is easy to put the atom in a corresponding position following the translation symmetry. For the quasicrystals, e.g., with well-known Penrose tiling, when an atom moves out of a boundary of a cell, one may expect to put the atom along the other equivalent boundary following the treatment for crystals. However, some of the edges usually belong to two different types of cells, whose atomic structures are independent in our model. Therefore, the approach is not acceptable. Here, we adopt another approach, i.e., reflecting the atom elastically by the edges/faces when it travels out of the cell. This approach is also widely applied in classical optimization.

In the implementation of the quasiOPT, we adopt the widely used embedded-atom method (EAM) following Zhou's formulism[14], which has been demonstrated feasible for a wide range of metals, including the transition metals.

The d-Al-Co-Ni quasicrystals have a series of thermodynamically stable phases over a broad range of compositions and temperatures.[15, 16] These two dimensional quasicrystals have periods along layered structures. Among the d-Al-Co-Ni quasicrystals, the so-called Ni-rich phase $Al_{72}Co_8Ni_{20}$ has attracted intensive studies due to its highly-ordered diffraction patterns. In our simulations, we employ a widely accepted two-layer model[16] with a period of about $c=4$Å along the z axis. The layers are generally denoted as $z=0$ and $z=c/2$. A framework containing only TM atoms is constructed according to Hiraga's HAADF-STEM images of $Al_{71}Co_{14.5}Ni_{14.5}$[17] (TM atoms can be recognized), as shown in Fig. 3. This configuration employs a Penrose tiling quasilattice, and its vertices with black and white dots correspond to different layers. All the vertices actually have similar local atomic structure as shown by two dashed circles in Fig. 3. It is clear that the basic cells of a given type can be treated identically.

Additional Al atoms should be filled into the above framework of the d-Al-Co-Ni quasicrystal since it is composed of about 70% Al atoms and 30% Co and Ni atoms. Experimentally, different diffraction patterns are found for d-Al-Co-Ni quasicrystals with various Co/Ni ratios.[15, 16] Our calculations indicate that the total energies of $Al_{147}Co_{23}Ni_{23}$ with various

Co/Ni distribution have differences within $6 \times 10^{-4}$ eV/atom. So for simplicity, we treat all the TM atoms as Ni atoms following the earlier theoretical approaches[6]. We conduct a quasiOPT simulation on a complete configuration based on the Yan-model[6]. The atomic structures around the vertex (the so-called 2-nm cluster) before and after relaxation are shown in Fig. 4a. We find that after relaxation, the inner two ring atomic structure distorts with respect to the third ring. The Al atom on the third ring relaxes a little inwards, implying an interaction with an atom on the second ring (see the two atoms indicated by arrows in Fig. $4a_4$). This indicates that atoms in the inner two rings are loose. In addition, it is clear that the region within the dashed line on z=0 layer is rather spacious for atoms and thus more atoms are expected to fill in to stabilize the quasicrystal. .

An improved configuration then is proposed, where some Al atoms are filled between the second and the third ring since there is much space according to the Yan-model based configuration. As expected, the improved configuration is energetically more stable (a formation energy reduction of 0.070 eV/atom). However, the Al atom on the third ring has an interaction with an Al atom outside this time (see the two atoms indicated by arrows), so it moves outward. The distortion implies that further improvement is required for the atomic configuration.

To fill new atoms into the spacious sites as in the improved model relies much on experience. Fortunately, with numerous simulations of d-Al-Co-Ni under various conditions, we find that the quasicrystal energetically prefers to form a special two-layer tiling, which obeys "rectangle-triangle" rules: i) One of the two layers is entirely constructed by two elementary cells called "triangle" and "rectangle" (cf. as the purple shapes in Fig. 5a). ii) The coupling between the two layers ensures that each vertex locates at a specific position of a cell on the other layer, i.e., for the triangle cell, the position has a same distance to the three vertices, while for the rectangle cell, the position has different distances to the terminals of the long edges, whose ratio is the golden ratio $\tau$, as the green dots shown in Fig. 5a. Although the rule of coupling between the layers is firstly suggested in this work, the similar rectangle-triangle tiling was mentioned in earlier literature[18]. When each vertex of the above tiling is occupied by an atom, the two rules guarantee that the bond lengths and angles are identical for all the nearest atoms, and can possibly maintain the fivefold symmetry. For instance, if we assume the atoms on the same layer have a nearest distance of 2.7Å, and the two layers are 2Å apart, the atoms on different layers then have a nearest distance of about 2.6Å, which coincides with d-Al-Co-Ni. In addition, this tiling has a maximum coordination number compared with previous structures with fivefold symmetry, indicating a potential better model.

Fig. 5b shows a cluster structure obeying the tiling rules as example. Its first layer (purple) maintains an ideal fivefold symmetry, while the second layer (green) does not as its atomic arrangement results from the tiling rules. For instance, the five atoms at the central region of the second layer (marked by arrows) cannot satisfy a fivefold symmetry, as three of them outside of the central regular pentagon while the other two just inside the pentagon. The broken symmetry at the central region propagates to the outer region. For instance, the non-equivalent angles shown in Fig. 5b also indicate that the fivefold symmetry is broken.

In fact, d-Al-Co-Ni quasicrystals do not always satisfy the ideal fivefold symmetry according to experimental observations, such as $Al_{72}Co_8Ni_{20}$[19, 20]. If we regard the five atoms connected by the black dashed line are TM atoms, the atomic configuration obtained from our special rules is consistent with Yan's Z-contrast image[20]. This indicates that our "rectangle-triangle" rules are consistent with the atomic configuration of d-Al-Co-Ni quasicrystals.

Following the above two configurations of d-Al-Co-Ni, we expect to develop a better configuration based on our "rectangle-triangle" rules. However, we find the "rectangle-triangle" rules are not completely compatible with Penrose tiling quasilattice due to their conflict over the fivefold symmetry. It is worthwhile to point out that the Penrose titling is not guaranteed to work perfectly for d-Al-Co-Ni quasicrystals, though it is widely adopted. It is possible that there is an undiscovered titling compatible with our "rectangle-triangle" rules, and fits d-Al-Co-Ni quasicrystals well.

Here, we adjust the atoms around the vertices of the fat and skinny cells to keep the fivefold symmetry at the center of Fig. 4$c_3$, and extend the local atomic distributions to the whole fat and skinny cells following the "rectangle-triangle" rules, and finally obtain the rectangle-triangle configuration as shown in Fig. 4c. The new configuration has a larger coordination number, as well as a much smaller relaxation in the quasiOPT simulation, in comparison with the other two configurations in Fig. 4, although some irregular pentagon appears at the central region of the cells (cf. the dashed lines in Fig. 4c).

The EAM calculated lattice constants and formation energies of three mentioned configurations of d-Al-Co-Ni, together with some Al-Ni alloys, are listed in Table 1. It is clear that the formation energy of d-Al-Co-Ni is comparable to most of the Al-Ni alloys. In particular, the formation energy of the "rectangle-triangle" configuration is lowered by 94 and 24 meV/atom in comparison with the Yan-model based and improved configurations, respectively. We notice that the stoichiometry of the "rectangle-triangle" configuration is deviated from $Al:TM \approx 30:70$ of the corresponding experimental sample. In fact, in the HAADF-STEM images, the brightness of the dots are clearly different[17], indicating that there are probably some bright dots are occupied by both Al and TM atoms with specific probability. We establish another "rectangle-triangle" configuration $Al_{137}TM_{56}$ by changing some Al atoms to TM atoms at some potential sites, with corresponding EAM calculated formation energy of 0.075 eV/atom. It is still much lower than that of Yan-model based configuration although their stoichiometry is similar to each other. Of note, although the positive value of formation energies for the alloys and d-Al-Co-Ni seems unphysical due to limitation of many EAM approaches[21, 22], it still implies the stability of the "rectangle-triangle" configuration.

A unique quasiOPT method, together with particular quasiperiodic boundary model, has been developed for the determination of detailed atomic structure of quasicrystals. Within the framework of a Penrose tiling quasilattice, a new configuration of d-Al-Co-Ni with lower formation energy is obtained by quasiOPT approach. In particular, a special two-layer tiling, whose layers are constructed by "triangle" and "rectangle" tiles with a proper coupling, is distilled for the d-Al-Co-Ni quasicrystals. This indicates that quasiOPT is efficient for atomic structure model description of quasicrystals.

**ACKNOWLEDGMENTS**

This work is supported by NSFC (Grant No. 11174082).

# Reference

[1]   A. Yamamoto and K. Hiraga, Phys. Rev. B **37**, 6207 (1988).
[2]   A. Yamamoto, K. Kato, T. Shibuya, and S. Takeuchi, Phys. Rev. Lett. **65**, 1603 (1990).
[3]   K. N. Ishihara and A. Yamamoto, Acta. Cryst. A **44**, 508 (1988).
[4]   E. Abe, Chem. Soc. Rev. **41**, 6787 (2012).
[5]   Y. Yan, S. J. Pennycook, and A. P. Tsai, Phys. Rev. Lett. **81**, 5145 (1998).
[6]   Y. F. Yan and S. J. Pennycook, Phys. Rev. Lett. **86**, 1542 (2001).
[7]   E. Abe, K. Saitoh, H. Takakura, A. P. Tsai, P. J. Steinhardt, and H. C. Jeong, Phys. Rev. Lett. **84**, 4609 (2000).
[8]   T. Tei-Ohkawa, K. Edagawa, and S. Takeuchi, J. Non-Cryst. Solids **189**, 25 (1995).
[9]   M. Mihalkovic, I. Al-Lehyani, E. Cockayne, C. L. Henley, N. Moghadam, J. A. Moriarty, Y. Wang, and M. Widom, Phys. Rev. B **65**, 104205 (2002).
[10]  P. H. Chen, K. Avchachov, K. Nordlund, and K. Pussi, Journal of Chemical Physics **138**, 234505 (2013).
[11]  D. Levine and P. J. Steinhardt, Phys. Rev. B **34**, 596 (1986).
[12]  J. E. S. Socolar and P. J. Steinhardt, Phys. Rev. B **34**, 617 (1986).
[13]  N. G. de Bruijn, Indagationes Mathematicae (Proceedings) **84**, 27 (1981).
[14]  X. W. Zhou, R. A. Johnson, and H. N. G. Wadley, Phys. Rev. B **69**, 144113 (2004).
[15]  S. Ritsch, C. Beeli, H. U. Nissen, T. Godecke, M. Scheffer, and R. Luck, Philosophical Magazine Letters **78**, 67 (1998).
[16]  K. Hiraga, T. Ohsuna, W. Sun, and K. Sugiyama, Materials Transactions **42**, 2354 (2001).
[17]  K. Hiraga and A. Yasuhara, Materials Transactions **54**, 493 (2013).
[18]  E. Cockayne, Phys. Rev. B **51**, 14958 (1995).
[19]  P. J. Steinhardt, H. C. Jeong, K. Saitoh, M. Tanaka, E. Abe, and A. P. Tsai, Nature **396**, 55 (1998).
[20]  Y. F. Yan and S. J. Pennycook, Nature **403**, 266 (2000).
[21]  F. Teng, W. Li, and Q. Yu, Adv. Mater. Res. **750-752**, 579 (2013).
[22]  Y. Qi, L. Wang, and T. Fang, Phys. Chem. Liq. **51**, 687 (2013).

Table 1. EAM calculated lattice constants and formation energies (eV/atom) of Al-Ni alloys and three configurations of d-Al-Co-Ni.

| Structure | Lattice (Å) | $N_{Al}$ | $N_{Ni}$ | $E_{Al}$ (eV) | $E_{Ni}$ (eV) | $\Delta H$ (eV) |
| --- | --- | --- | --- | --- | --- | --- |
| Al1Ni1 | 2.987 | 1 | 1 | -3.993 | -3.902 | 0.068 |
| Al1Ni3 | 3.646 | 1 | 3 | -4.109 | -4.229 | 0.034 |
| Al3Ni2 | 4.203, 5.142 | 3 | 2 | -3.827 | -3.638 | 0.177 |
| Al3Ni5 | 7.748, 6.957, 3.874 | 6 | 10 | -4.026 | -4.123 | 0.037 |
| Al3Ni1 | 6.920, 7.710, 5.035 | 12 | 4 | -3.745 | -3.552 | 0.101 |
| Al4Ni3 | 11.913 | 64 | 48 | -3.870 | -3.772 | 0.125 |
| Yan-model based | 21.0, 4.16 | 118 | 46 | -3.682 | -3.600 | 0.165 |
| Improved | 21.2, 4.32 | 135 | 46 | -3.707 | -3.705 | 0.095 |
| Rectangle-triangle | 21.5, 4.39 | 147 | 46 | -3.697 | -3.778 | 0.071 |

**Figure captions**

Fig. 1 A simple quasicrystal (c) with fivefold symmetry is described by (a) quasilattice and (b) basic cells. In (c), the basic cells of a specific type have the same atomic structure as illustrated in (b).

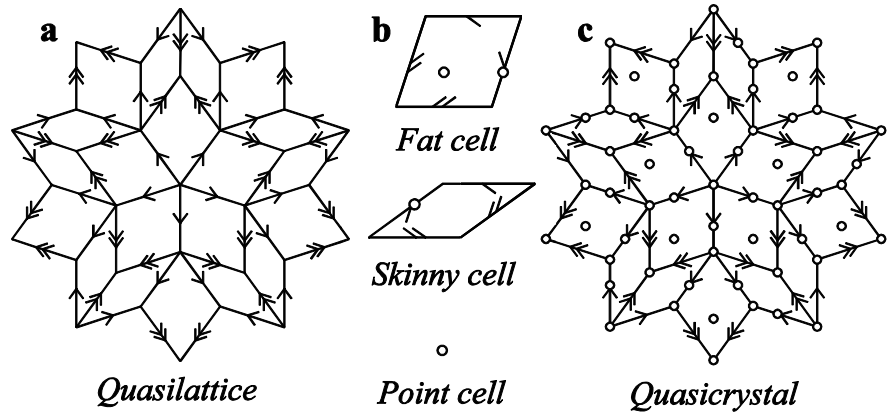

Fig. 1 Xiao-Tian Li, et al.

Fig. 2 (Color online) The relaxing region (in magenta) of quasiOPT contains (a) a complete set of basic cells and (b) basic cells with all the possible neighboring arrangement of cells. The boundary regions are shown in cyan, and the black dots in (b) correspond to the eight types of vertices in the Penrose tiling.

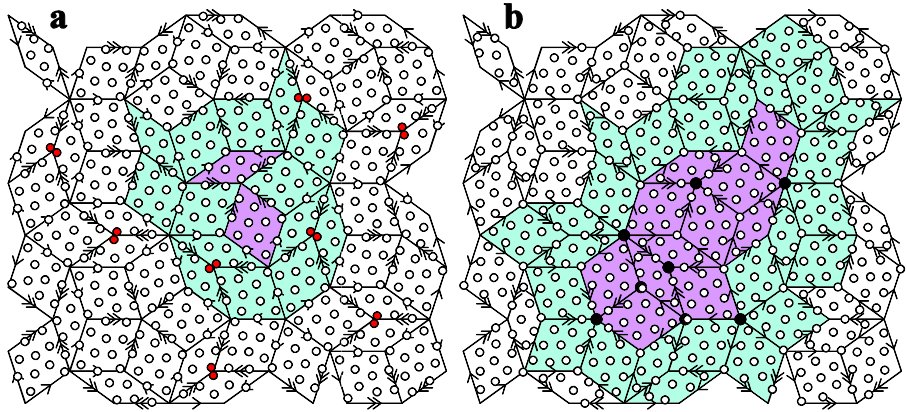

Fig. 2 Xiao-Tian Li, et al.

Fig. 3 (Color online) A framework containing only TM atoms with Penrose tiling quasilattice. The relaxing and boundary regions are shown in magenta and cyan, respectively. The purple and green circles denote the TM atoms on z=0 and z=c/2 layers, respectively, and the black and white dots correspond to two different types of vertices.

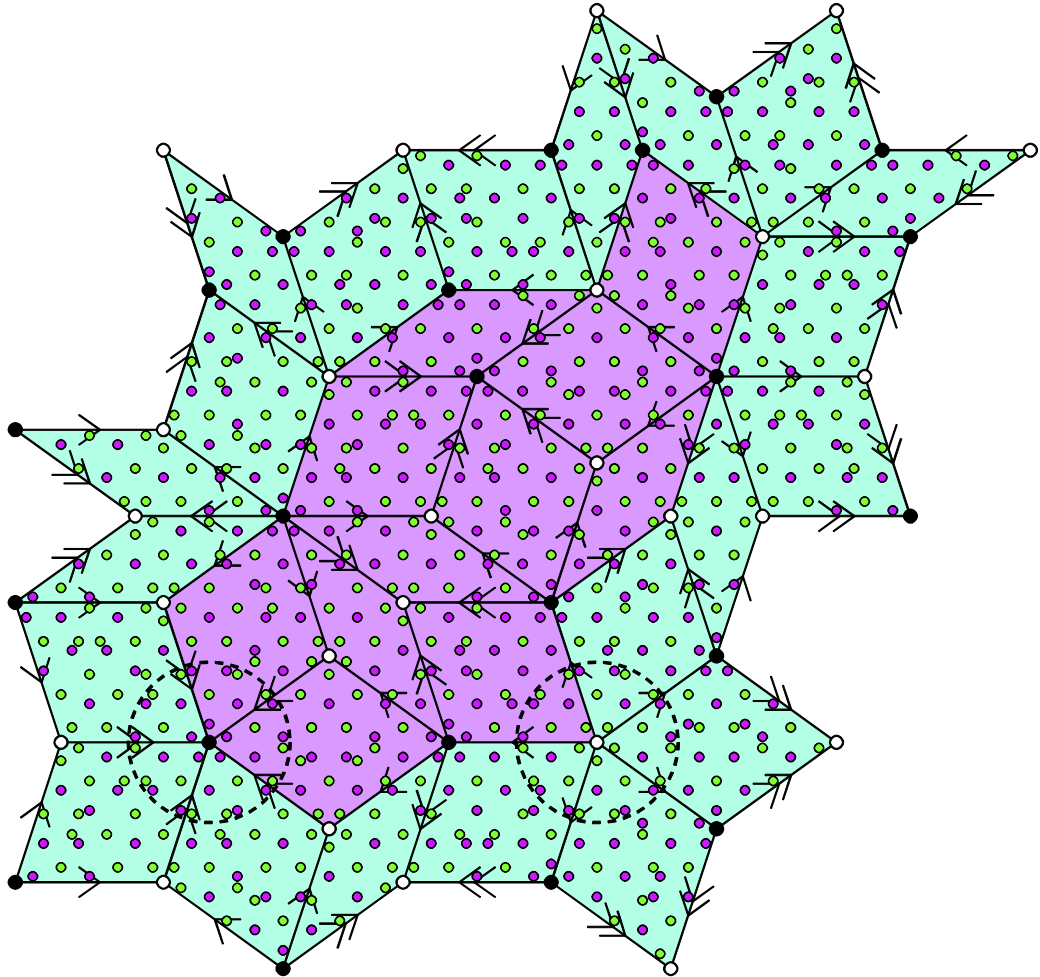

Fig. 3 Xiao-Tian Li, et al.

Fig. 4 (Color online) The atomic structure of Yan-model based configuration, improved configuration, and rectangle-triangle configuration. Besides the fat and skinny cells, there is a third cell containing only an Al atom at each vertex, which is not illustrated here (see the atoms at the center of the cluster). The clusters are the atomic structures around a specific vertex, as indicated by the dash circle in the lower left corner of Fig. 3. Atoms on z=0 and z=c/2 layers are denoted by purple and green circles, respectively, while large and small circles are corresponding to the TM and Al atoms. The black and white dots represent two different types of vertices.

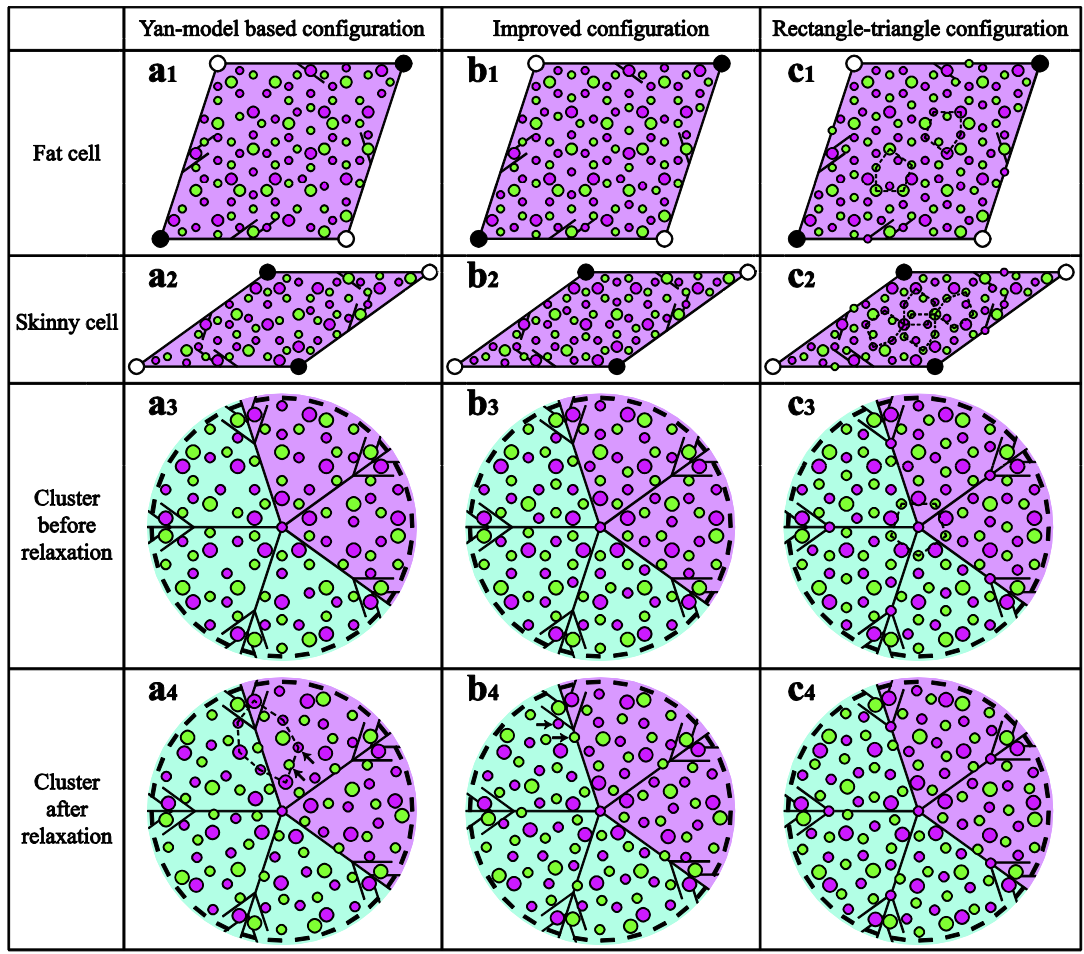

Fig. 4 Xiao-Tian Li, et al.

Fig. 5 (Color online) Illustration of the "rectangle-triangle" rules with (a) two elemental tiles of a two-layer tiling and (b) a cluster structure constructed from the two elemental tiles shown in (a).

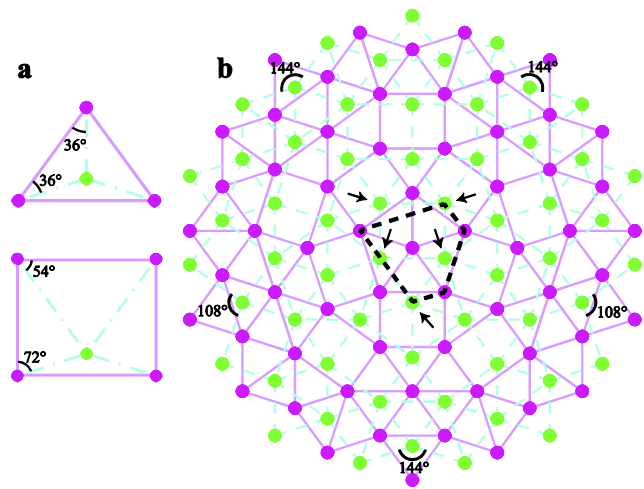

Fig. 5 Xiao-Tian Li,et al.